\documentclass[aip,rsi,twocolumn,reprint,amsmath,amssymb,superscriptaddress,longbibliography,floatfix]{revtex4-2}%

\usepackage{graphicx}% Include figure files
\usepackage{dcolumn}% Align table columns on decimal point
\usepackage{bm}

\usepackage{color}

\usepackage{mathptmx}
\usepackage{mathrsfs}
\usepackage{multirow}
\usepackage{natbib}

\usepackage{hyperref}
\usepackage{epsf}
\usepackage{epsfig}
\usepackage{epstopdf}
\newcommand{\MN}[2]{{\color{black}#2}}                %just text-without comments

\begin{document}

\title{\MN{}{Heat-capacity measurements under uniaxial pressure using a piezo-driven device}}

\author{Y.-S.\ Li}
\affiliation{Max Planck Institute for Chemical Physics of Solids,  N\"{o}thnitzer Str.\ 40, 01187 Dresden, Germany}
\affiliation{Scottish Universities Physics Alliance \MN{}{(SUPA)}, School of Physics and Astronomy, University of St Andrews, St Andrews \MN{}{KY16 9SS}, \MN{}{United Kingdom}}

\author{R.\ Borth}
\affiliation{Max Planck Institute for Chemical Physics of Solids,  N\"{o}thnitzer Str.\ 40, 01187 Dresden, Germany}

\author{C.\ W.\ Hicks }
\affiliation{Max Planck Institute for Chemical Physics of Solids,  N\"{o}thnitzer Str.\ 40, 01187 Dresden, Germany}

\author{A.\ P.\ Mackenzie }
\affiliation{Max Planck Institute for Chemical Physics of Solids,  N\"{o}thnitzer Str.\ 40, 01187 Dresden, Germany}
\affiliation{Scottish Universities Physics Alliance \MN{}{(SUPA)}, School of Physics and Astronomy, University of St Andrews, St Andrews \MN{}{KY16 9SS}, \MN{}{United Kingdom}}

\author{M.\ Nicklas}
\email{Michael.Nicklas@cpfs.mpg.de}
\affiliation{Max Planck Institute for Chemical Physics of Solids,  N\"{o}thnitzer Str.\ 40, 01187 Dresden, Germany}

\date{\today}

\begin{abstract}
We report the development of a technique to measure heat capacity at large uniaxial pressure using a piezoelectric-driven device generating compressive and tensile strain in the sample. Our setup is optimized for temperatures ranging from 8~K down to millikelvin. Using an AC heat-capacity technique we are able to achieve an extremely high resolution and to probe a homogeneously strained part of the sample. We demonstrate the capabilities of our setup on the unconventional superconductor Sr$_2$RuO$_4$. By replacing thermometer and adjusting the remaining setup accordingly the temperature regime of the experiment can be adapted to other temperature ranges of interest.

\end{abstract}

\maketitle

\section{INTRODUCTION}

Application of external pressure is a powerful method to tune the intricate interplay of competing energy scales in correlated materials and the emergence of novel unconventional phases in a clean fashion. It offers significant advantages as a control parameter compared with chemical substitution and application of magnetic fields, because it does not introduce additional disorder as in the case of substitution of one element by an other or polarize the electrons as a magnetic field does.

A large variety of experimental setups has been developed to probe physical properties under hydrostatic pressure.\cite{Nicklas2015} In contrast, experiments under uniaxial pressure appeared to be limited to low pressure and only few experimental probes. Recently, however, the development of piezoelectric-driven pressure devices opened a new perspective.\cite{Hicks2014a,Barber2019} These devices allow the application of large positive and negative pressures and the amplitude of the applied pressure can be easily changed at low temperatures. In a short period of time experimental stages to access a large number of physical properties of materials have been developed. These include electrical transport,\cite{Stern2017,Steppke2017} magnetic susceptibility,\cite{Hicks2014b} nuclear-magnetic resonance,\cite{Kissikov2017,Luo2019,Pustogow2019} muon-spin resonance,\cite{Grinenko2020} and angle-resolved photo emission, for which mechanically or thermally activated cells have also been introduced.\cite{Flototto2018,Ricco2018,Pfau2019a,Pfau2019b,Sunko2019}

An important quantity to characterize a material is the specific heat, which is the fundamental thermodynamic property giving information on the internal degrees of freedom of a material and the entropy related with them.
To address the experimental challenge of studying the heat capacity under large uniaxial pressures, we employ a variation of known AC heat-capacity measurement techniques.\cite{Sullivan1968} Heat capacity measurement has been combined with uniaxial pressure previously,\cite{Jin1992,Reinders1994,Miclea2002,Zieve2004,Dix2009} but with traditional, anvil-based uniaxial pressure cells. Samples have been thermally isolated by using low thermal conductivity materials, such as stainless steel or superconducting NbTi, as piston or additional spacer.  However in previous anvil-based uniaxial-pressure measurements, e.g.\ on the unconventional superconductor Sr$_2$RuO$_4$, it did not prove practical to maintain high stress homogeneity,\cite{Kittaka2010,Taniguchi2015} which is one of the main challenges in carrying out this kind of experiments. Furthermore, under applied uniaxial pressure the samples even may deform plastically. To reduce these effects we apply force to the sample through a layer of epoxy,\cite{Hicks2014a} which acts as a conformal layer that dramatically improves stress homogeneity. However, it also makes heat-capacity measurement more challenging, because the epoxy layer provides an unavoidable strong thermal link to the pressure cell.

For our study we have used Sr$_2$RuO$_4$ that provides a demanding test of our new apparatus. Sr$_2$RuO$_4$ is an unconventional superconductor with a superconducting transition temperature up to $T_c=1.5$~K in the best crystals.\cite{Mackenzie1998,Mackenzie2003,Mackenzie2017} From resistivity and magnetic susceptibility experiments it is known that $T_c$ shows a pronounced dependence on the applied uniaxial pressure.\cite{Hicks2014b,Steppke2017,Barber2018} This and the sharp superconducting transition anomaly make it an ideal material to demonstrate the potential of our technique for the study of correlated materials. A successful experiment on Sr$_2$RuO$_4$ can only be done using a technique that introduces no disorder or plastic deformations, and that probes a region in which the strain induced in the sample is highly homogeneous.

\section{METHOD}

For a setup in which the sample is strongly coupled to the environment as it is in a pressure cell, whether hydrostatic or uniaxial, standard quasi-adiabatic or relaxation techniques are limited to cases where the heat capacity of the whole pressure cell including the sample is measured and the heat capacity of the sample can then be separated from the (large) addenda. That implies restrictions on the materials which can be investigated and limits experiments to low temperatures. The advantage of such a technique is that one obtains absolute values of the heat capacity, but the resolution and the pressure regime are limited. In heat-capacity measurements at higher pressure, where anvil-type cells are used, or for uniaxial pressure experiments, the application of this technique is not possible anymore. The mass of sample is negligible with respect to that of the pressure apparatus. In these cases the heat capacity can only be determined using an AC heat capacity measurement technique.\cite{Sullivan1968} With the AC technique it is possible to record heat capacity data in a wide range of parameter space on a sample which is not well thermally isolated from its environment by adjusting the measurement frequency. The drawback is that it is generally challenging to obtain absolute values of the heat capacity and one usually has to be content with data having arbitrary units. As we demonstrate, however, it can still yield a wealth of useful information.

\subsection{AC Heat Capacity}\label{AC Heat Capacity}

\begin{figure}[tb!]
\includegraphics[width=0.9\linewidth]{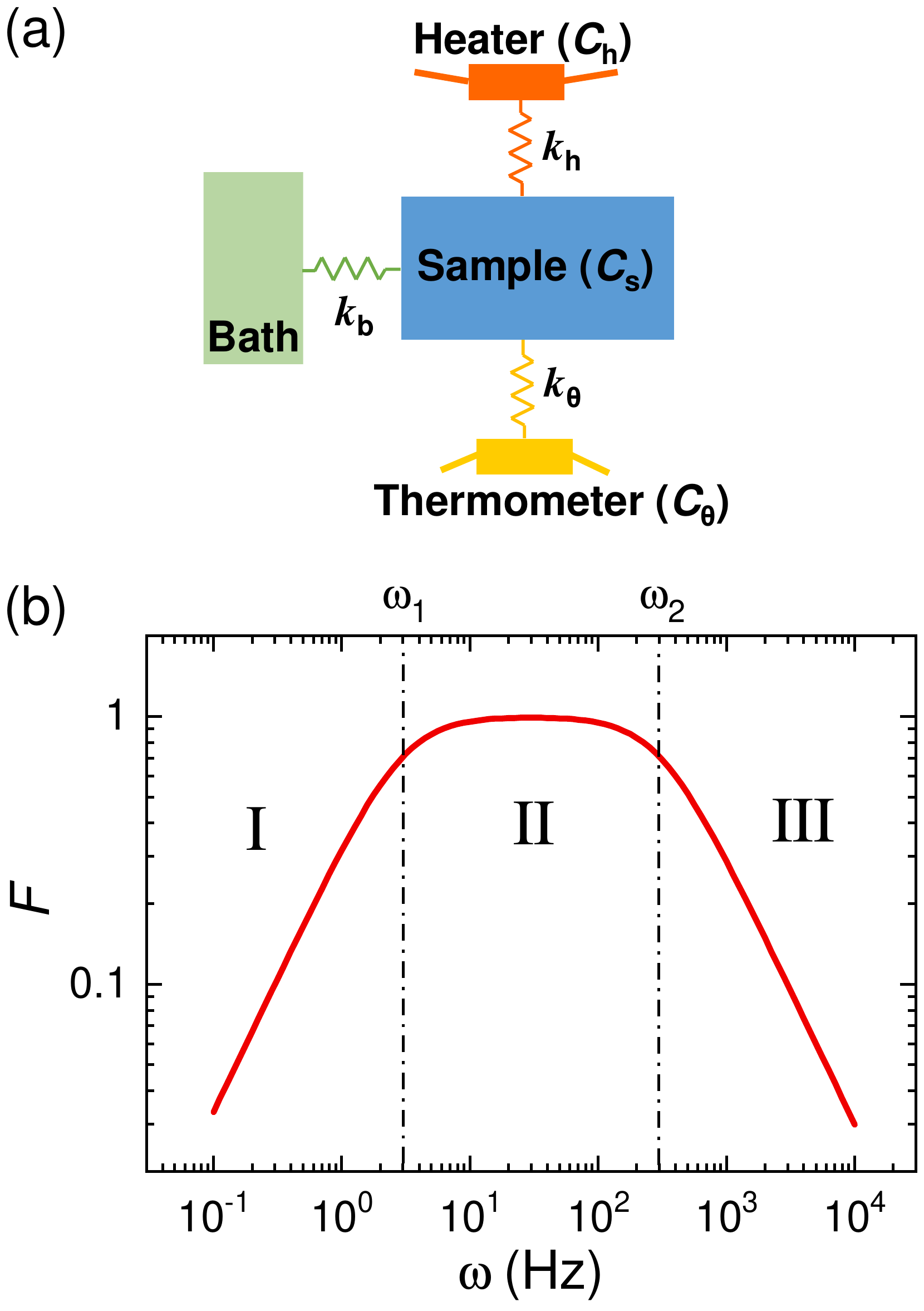}
\centering
\caption{
(a) Schematic drawing of the thermal couplings of a sample in an AC heat-capacity setup.
(b) A schematic diagram of the frequency response curve $F$ against $\omega$. $\omega$ is the angular frequency of the temperature oscillation. The curve can be divided into three regions separated by $\omega_1$ and $\omega_2$.
 }
\label{Scheme_HC}
\end{figure}

In the AC heat capacity measurement technique an alternating current is applied at frequency $\omega/2$ to the heater, leading to an AC heat power at frequency $\omega$ to determine the heat capacity $C_{AC}$. Here $\omega=2\pi f$ is the angular frequency.
The governing relationship for measurements of the AC heat capacity is
\begin{equation}
  C_{AC}=  \frac{P}{\omega T_{AC} } F(\omega).
\label{Cac}
\end{equation}
$P$ is the average power and $F(\omega)$ is a frequency response curve that characterizes the thermalization of the sample, and differs from sample to sample, because it depends on time constants determined by thermal conductances and heat capacities of the system.
$F(\omega)$ depends on the time constants $\tau_1$ and $\tau_2$:
\begin{equation}\label{F_omega}
  F(\omega)=\left[ 1+\frac{1}{\omega^2\tau_1^2}+\omega^2\tau_2^2\right] ^{-1/2}.
\end{equation}
$\tau_1=C_{AC}/k_b$ describes the time scale of the applied heat power decaying to the environment, whereas  $\tau_2=\sqrt{\tau_h^2+\tau_{\theta}^2+\tau_{\rm int}^2}$ describes the internal thermal time scale within the system itself.
Here $\tau_h=C_h/k_h$,$\tau_{\theta}=C_{\theta}/k_{\theta}$  and $\tau_{\rm int}=C_s/k_s$ (see also Fig.\ \ref{Scheme_HC}a). The time constants $\tau_h$ $\tau_{\theta}$ and $\tau_{\rm int}$ describe the time scales for the heater and, thermometer and sample to be thermalized, respectively. $C_h$, $C_{\theta}$, and $C_s$ are the heat capacity of the heater, thermometer, and sample, respectively. For a good design, the responses of heater and thermometer need to be fast so one should aim at $\tau_{\rm int}\gg \tau_h$ and $\tau_{\theta}$.

A schematic diagram  of $F(\omega)$ is shown in Fig.\ \ref{Scheme_HC}b. At low frequencies, indicated as regime I, $\omega\ll\omega_1= 1/\tau_1=k_b/C_{AC}$, $F(\omega)$ is reduced due to dissipation of temperature oscillations into the environment and at high frequencies, $\omega\gg\omega_2= 1/\tau_2\cong k_s/C_s$, because the heater-sample-temperature sensor system does not thermalize, marked as regime III.
In the plateau region between these limits $F(\omega)\approx 1$ and
\begin{equation}\label{C_ac_prox}
C_{AC}\approx  \frac{P}{\omega T_{AC} }.
\end{equation}

In addition to the temperature oscillations the application of the oscillatory heating power leads to an temperature offset $T_{DC}$ in the sample, which can be determined in the low frequency limit $\omega\ll\omega_1$. Here $F(\omega)=\omega\tau_1$ and the temperature offset can be estimated as
\begin{equation}\label{T_DC}
T_{DC}\approx  \frac{P}{k_b}.
\end{equation}

\subsection{Experimental Setup}

\begin{figure}[tb!]
\includegraphics[width=0.9\linewidth]{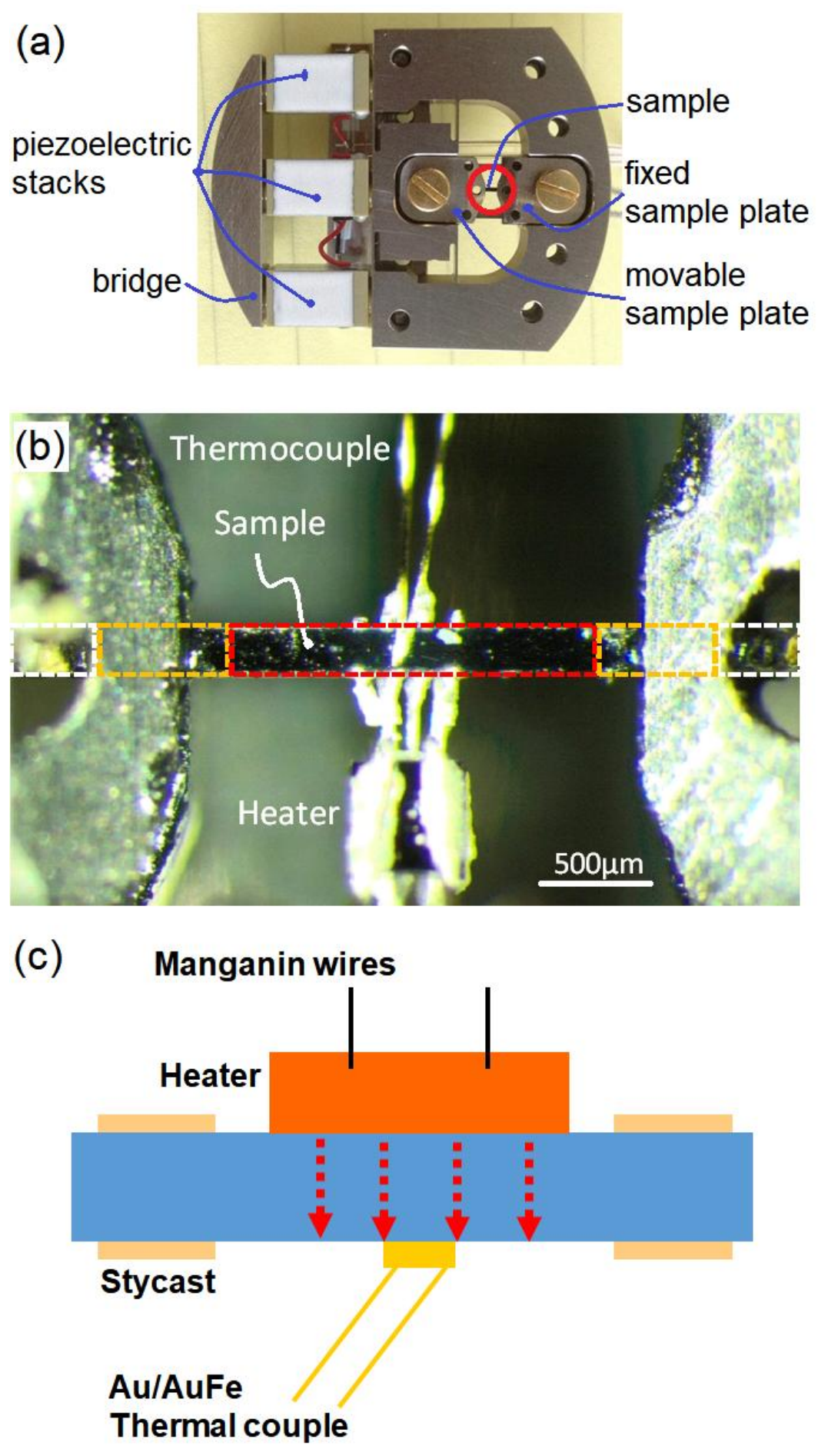}
\centering
\caption{
(a) Photograph of the uniaxial pressure apparatus used in the present study. The red circle marks the sample region.
(b) Photograph of the setup of the heat capacity measurements under strain including heater and thermometer. The sample is glued between the jaws of the uniaxial pressure device.  \MN{}{The exposed length, width and thickness of the shown sample are 2~mm, 200~$\mu$m and 150~$\mu$m, respectively. The device} allows the application of compressive and tensile strains. The red, yellow, and white rectangles represent the (quasi)homogeneous, inhomogeneous, and unstrained regions, respectively, see text for details.
(c) Schematic diagram of the setup illustrating the photograph in (b).}
\label{Scheme_thermalConductivity}
\end{figure}

The general considerations in the section above show that it should be possible to measure the heat capacity in an uniaxial pressure apparatus as shown in Fig.\ \ref{Scheme_thermalConductivity}a by choosing the correct set of experimental parameters. In the following we will explain the experimental setup and describe the details of the preparation process using the example of a Sr$_2$RuO$_4$ single crystal.

The sample is marked by a red circle in Fig.\ \ref{Scheme_thermalConductivity}a and shown in detail in Fig.\ \ref{Scheme_thermalConductivity}b. In this setup the applied force results in a normal strain
\begin{equation}\label{strain}
\varepsilon_{xx}=\frac{l-l_0}{l_0}
\end{equation}
in the sample. Here $l_0$ is the length of the unstrained sample and $l$ the length of the strained sample. The length change is measured capacitively and can be controlled. \MN{}{The applied strain can go beyond $1\%$. In Sr$_2$RuO$_4$ the Young's modulus is about 180~GPa and correspondingly the applied uniaxial pressure can reliably reach up to about 2~GPa. However, the maximum uniaxial pressure depends strongly on the mechanical properties of the investigated material.} Further details can be found in Ref.\ \onlinecite{Hicks2014a}.
The present AC heat-capacity technique can be adapted to different types of uniaxial pressure devices, e.g.\ to a stress-controlled apparatus \cite{Barber2019} \MN{}{and  is fully compatible with experiments in magnetic fields}

In Fig.\ \ref{Scheme_thermalConductivity}b we show a photograph of the bar-shaped sample that has been carefully cut, polished, and then mounted within the jaws of the uniaxial pressure rig.
The nature of the apparatus means that only the central part of the sample is homogeneously strained. Force is transferred to the sample through the epoxy layer around the sample. The sample ends which are protruding beyond it are unstrained, and there are intermediate regions, marked in yellow in Fig.\ \ref{Scheme_thermalConductivity}b, where the strain is built up. Therefore, we have to choose the measurement conditions in a way that we only probe the homogeneous part of the sample. On the example of a Sr$_2$RuO$_4$ single crystal we will demonstrate that this is in principle possible by varying the excitation frequency $f_{\rm exc}=f/2$ of the heater, if the characteristic parameters of the setup, such as the different thermal conductances, have been chosen in the appropriate range.

For the experiments single crystalline Sr$_2$RuO$_4$ was aligned using a bespoke Laue x-ray camera, and cut using a wire saw into thin bars with whose long axis aligned with the [100] direction of the crystal. For the best results these bars were polished using home-made apparatus based on diamond impregnated paper with a minimum grit size of 1~$\mu$m. The bar was then mounted within the jaws of the uniaxial pressure rig using Stycast 2850FT epoxy with Catalyst 23LV (Henkel Loctide). A resistive thin film resistor chip (State of the Art, Inc., Series No.:\ S0202DS1001FKW) as heater and a Au-AuFe(0.07\%) thermocouple are fixed to opposite sides of the sample using Dupont 6838 single component silver-filled epoxy. \MN{}{The resistance of heater is about 640~$\Omega$ and the applied power is in the range of $\mu$W. The heater is connected electrically using manganin wires providing a low thermal conductance to the bath. At 1~K the thermal conductance of the Stycast layers and the manganin wires is about $10^{-4}$ and $10^{-7}$~ W/K, respectively. Thus, the heat loss is largely dominated by the Stycast layers.} The thermocouple was spot-welded in-house and its calibration fixed by reference to that of a calibrated RuO$_2$ thermometer.\footnote{In the temperature range between 0.15 and 4.5~K the thermopower $S$ of the Au-AuFe(0.07\%) thermocouple is described by $S(T)=[10.1483\cdot T/{\rm K}-8.75772\cdot (T/{\rm K})^2+4.00231\cdot (T/{\rm K})^3-0.838741\cdot (T/{\rm K})^4+0.0667604\cdot (T/{\rm K})^5]{\rm ~\mu V/K}$ .} Special care was taken when epoxying to the pressure cell to minimize tilt and ensure an as homogeneous strain field as possible.

The uniaxial pressure apparatus was mounted on a dilution refrigerator (Oxford Instruments), with thermal coupling to the mixing chamber via a high purity silver wire. The data were acquired between 500~mK and 4.2~K, with operation above 1.5~K achieved by circulating a small fraction of the mixture.  The extremely low noise level of 20~${\rm pV/\sqrt{Hz}}$ on the thermocouple readout was achieved by the combination of an EG\&G 7265 lock-in amplifier and a \MN{}{high frequency} low temperature transformer (\MN{}{LTT-h from} CMR direct) mounted on the 1~K pot of the dilution refrigerator, operating at a gain of 300. \MN{}{The input impedance of the transformer is about $0.1~\Omega$, which ensured a flat frequency response from several hundred Hz to several tens of kHz.} A Keithley 6221 low-noise current source was used to drive the heater. The piezo-electric actuators were driven at up to $\pm400$ V using a bespoke high-voltage amplifier.

\subsection{Strain inhomogeneity}

The nature of our setup is that the strain profile along the direction of the application of the uniaxial pressure is not homogeneous.
As we will describe below, by adjusting the excitation frequency $f_{exc}$ to an appropriate value the actual heat-capacity measurement can be confined to the quasi-homogeneously strained region of the sample. Besides this source of strain inhomogeneity there are other sources which can be reduced in the preparation and mounting process of the sample in the apparatus.

\subsubsection*{Imperfections of sample surface and geometry}

The bar-shaped needles cut from crystals have typically terraces and irregular shapes on their surface which can induce inhomogeneous strain fields when they are under uniaxial pressure. Imperfections may also lead to an early failure of pressurized samples reducing the maximum achievable pressure.
A perfect sample is a cuboid, i.e.\ each surface is parallel to the opposite one, and has a smooth surface roughness. Therefore, we carefully polish our samples and inspect the shape and the surface quality under a microscope before mounting in the uniaxial pressure apparatus.

\begin{figure}[tb!]
\includegraphics[width=0.95\linewidth]{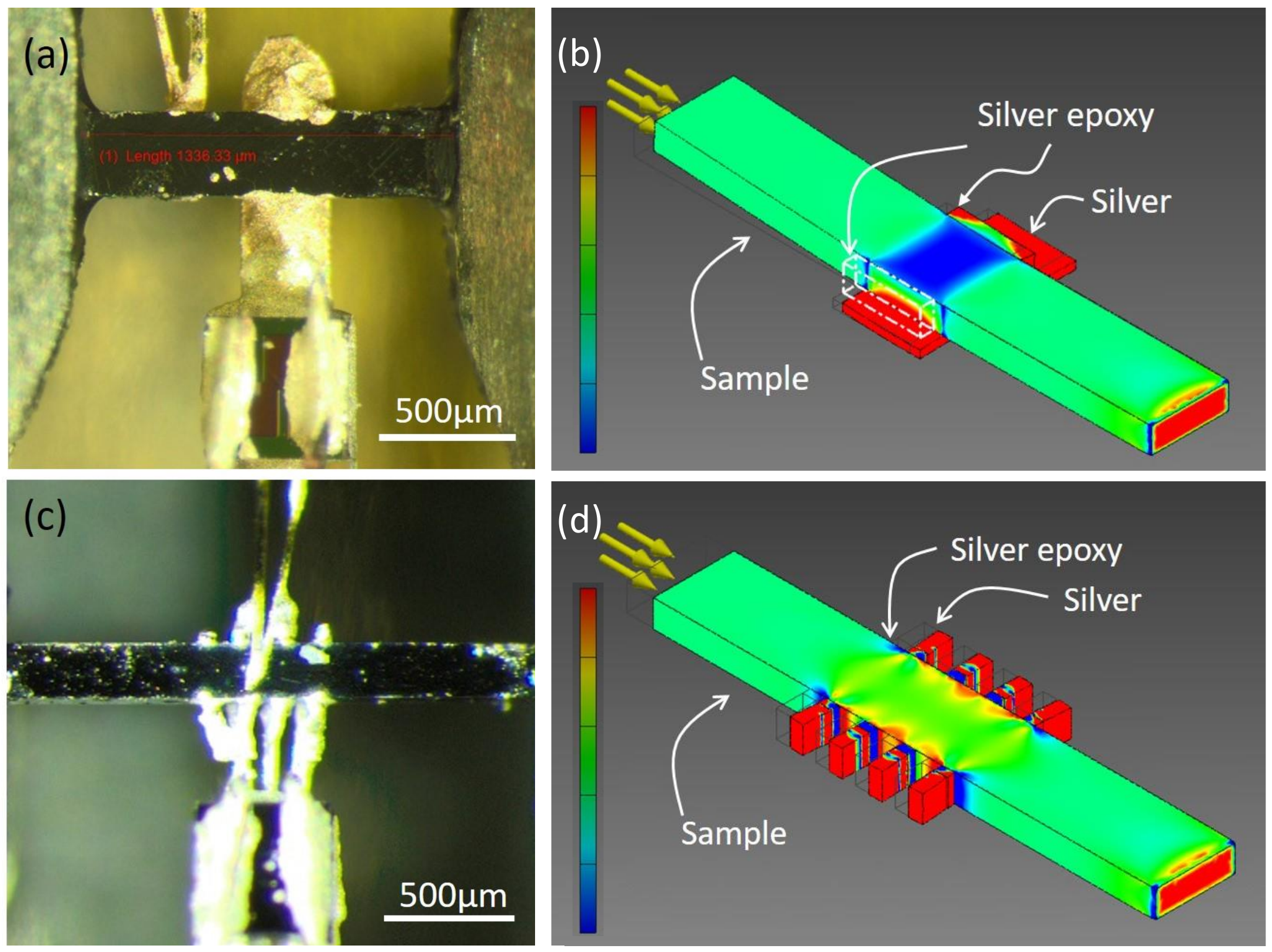}
\centering
\caption{(a) Heater fixed with a silver foil to the sample. The contact to the sample is on the whole plane.
(b) The simulation of the strain $\varepsilon_{xx}$ pattern corresponding to the setup in (a). One of the silver-filled epoxy blocks was set to be invisible, as indicated by the
dash-dotted lines, such that the strain proﬁle on the edge of the sample is visible.
(c) Heater fixed to the sample by four thin silver wires on the edges of the sample.
(d) Corresponding simulation to (c). The strain inhomogeneity is reduced in the center compared with the setup shown in (a).
 }
\label{heater_mounting}
\end{figure}

\subsubsection*{Bending}

Asymmetric mounting of a sample leads to bending.\cite{Barber2017} An ideal sample mounting is a sample mounted between two plates with symmetrical epoxy layers on top and at the bottom. However, the sample might end up with a small offset in height. To reduce inhomogeneity in preparing the sample we aim for an aspect ratio $l_s/t >10$,  where $l_s$ is the exposed length and $t$ the thickness of the sample

\subsubsection*{Mounting of the heater}

One of the main sources of inhomogeneous strain fields originates from the sample configuration in the AC heat-capacity setup. In order to transmit the heating power from the heater resistance to the sample, we use thermal contacts made by silver wires glued to the sample using silver-filled epoxy. Since the Young’s modulus of silver and the sample are generally very different, as in our example of Sr$_2$RuO$_4$, the contacts create inhomogeneous strain fields. We tried to minimize this effect. We realized two different types of silver contacts to the sample. Figures \ref{heater_mounting}a and \ref{heater_mounting}c show photographs of the setups. In the first one a silver strip was glued on a contact length of about 300~$\mu$m on both edges to the sample using silver-filled epoxy. In the second, the thermal contact is divided into 4 smaller areas instead of a large one, by gluing 8 silver wires with diameter of 50~$\mu$m on both edges. The total contact area in both cases is almost the same.
The experiments on our test sample Sr$_2$RuO$_4$ showed indeed a significant sharpening of the superconducting transition anomaly in the latter case.

In addition to the experiments we simulated the strain fields in the sample by a finite element method using a commercial software package.\footnote{Autodesk Inventor 2015, Autodesk Inc..}
For the simulation we set the Young’s modulus of the sample to 180 GPa and the Poisson’s ratio to 0.33. For the dimensions of the sample we used the values from the experiment, a thickness 100~$\mu$m, width 300~$\mu$m, and length 2~mm. One of the sample ends was set to be fixed and the other end was subjected to a pressure of 0.18~GPa, leading to $\varepsilon_{xx}=0.1$\%. The silver and silver-filled epoxy were set to have Poisson’s ratio of 0.35. The Young’s modulus for the silver epoxy was set to be $1/3$ of that of the silver, which is 110~GPa. The results for both configurations are shown in Figs.\ \ref{heater_mounting}b and \ref{heater_mounting}d. The color bar shows the strain scale, ranging from 0.07 to 0.13\%. In the first configuration, with silver strip glued with silver-filled epoxy on both edges to the sample, the strain inhomogeneity on the sample is greater than 60\% in the center (see Fig.\ \ref{heater_mounting}b).
In the second design using 8 silver wires with diameter of 50~$\mu$m on both edges for the thermal contact, the strain inhomogeneity is strongly reduced in the bulk, except in the regions very close to the contact surfaces. Since heat capacity is a bulk-sensitive measurement, the inhomogeneity near the surface is negligible. The strain inhomogeneity in this configuration is only about $10$\% in the center region of the sample. This shows that it is highly desirable to have separated smaller contact areas to transmit the heat to the sample in order to reduce strain inhomogeneities in accordance with the experimental results. In the following we continued with the second configuration.

\section{RESULTS}

We demonstrate the capabilities of our setup and discuss its advantages and limitations by showing representative data from experiments on Sr$_{2}$RuO$_{4}$. The first step in an AC heat-capacity experiment is to find a suitable measurement frequency in the plateau region of the frequency response curve $F(\omega)$. We note that the existence of this plateau depends on the respective characteristics of the setup as discussed in Sec.\ \ref{AC Heat Capacity}. If a suitable frequency has been found, temperature \MN{}{sweeps, also in applied magnetic field, or pressure/magnetic field ramps} can be conducted and the heat capacity recorded. According to Eq.\ \ref{C_ac_prox} we will plot our results on Sr$_{2}$RuO$_{4}$ as $P/[\omega T_{AC}(T)]$. As we will discuss in Sec.\ \ref{Cv} $C_{AC}\approx P/(\omega T_{AC})$ is not strictly valid in our setup and has to be treated with caution.

\subsection{Measuring frequency}

\begin{figure}[tb!]
\includegraphics[width=0.95\linewidth]{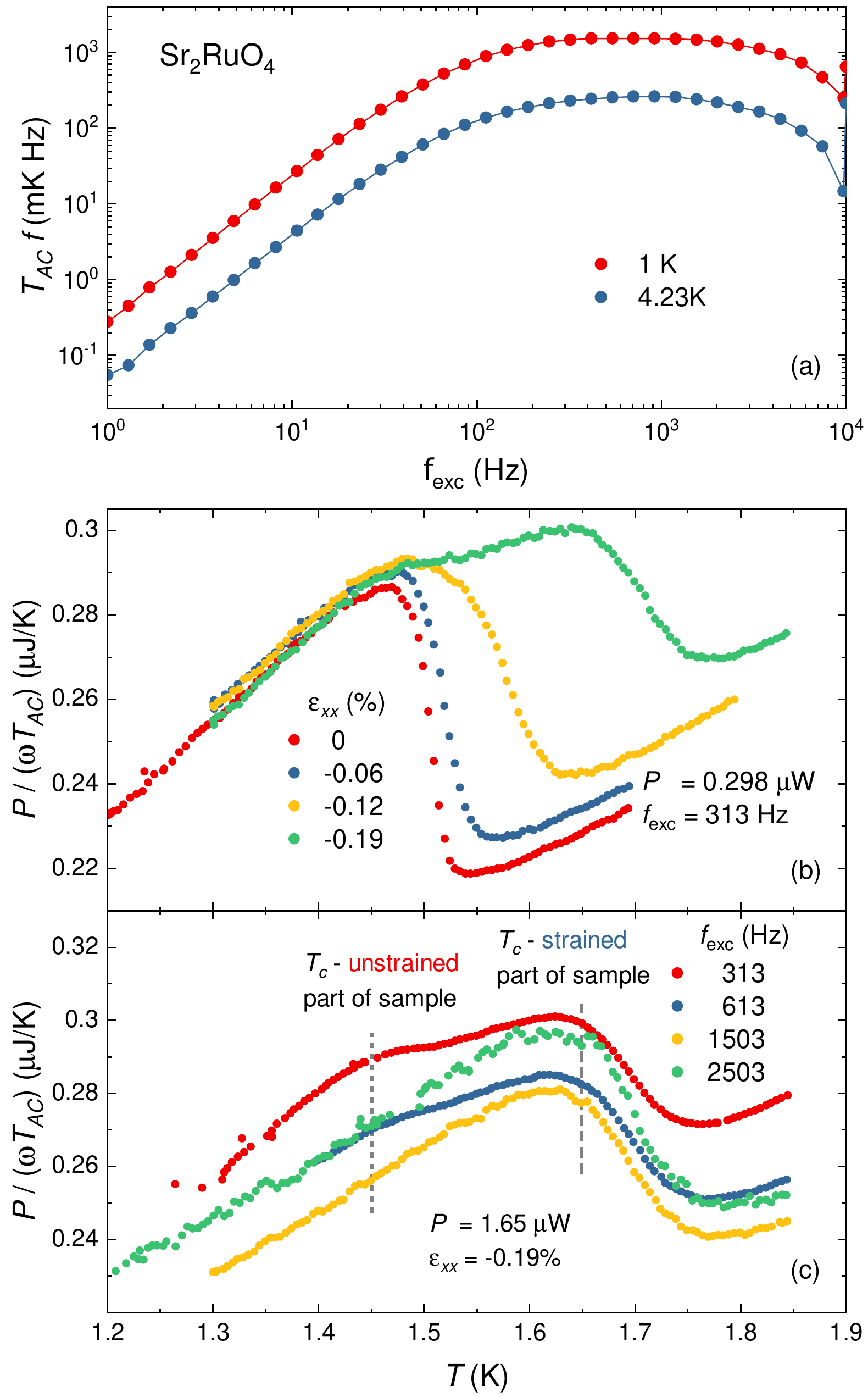}
\centering
\caption{(a) Frequency sweeps at 1 and 4.23~K.
(b) Data  recorded at 313~Hz for zero and small strains up to $\varepsilon_{xx}=-0.19$\%.
(c) Data on  Sr$_2$RuO$_4$ in the region around its superconducting transition at $\varepsilon_{xx}=-0.19$\% for different frequencies.
 }
\label{frequency_effect}
\end{figure}

Figure \ref{frequency_effect}a shows the frequency response at 1 and 4.23~K in case of our example Sr$_2$RuO$_4$ crystal. It shows a broad plateau between a few hundred hertz and several kilohertz at both temperatures, attesting that in principle heat-capacity measurements should be possible in the desired temperature range. By raising the temperature from 1 to 4.23~K the plateau narrows slightly but remains well-defined.

In the lower frequency part of the plateau in Fig.\ \ref{frequency_effect}a, temperature oscillations extend throughout the sample and all three regions the homogenously strained in the center, the unstrained portions at the ends and the regions where strain builds up are probed in a measurement (see Fig.\ \ref{Scheme_thermalConductivity}b). Figure \ref{frequency_effect}b shows $P/[\omega T_{AC}(T)]$ recorded at $f_{exc}=313$~Hz for different $\varepsilon_{xx}$. At zero strain we see a single sharp transition anomaly at $T_c\approx1.45$~K. Upon increasing $|\varepsilon_{xx}|$ the step-like feature moves to higher temperatures, consistent with the increase in $T_c$ with strain,\cite{Hicks2014b} but a second feature remains at the original zero-strain transition. This latter feature stems from the unstrained part of the sample.

To reduce the size of the probed part of the sample and restrict it to the homogenously strained region in the center, we increased the measurement frequency. We note that we still stay in the plateau region of frequency response curve. To demonstrate the importance of this increase in measurement frequency, we applied modest strain $\varepsilon_{xx}=-0.19$~\% and increased $f_{exc}$ from 313~Hz in steps to 2503~Hz. The data are displayed in Fig.\ \ref{frequency_effect}c.
At 313 and 613~Hz, in addition to the peak at $\approx 1.65$~K corresponding to the transition in the central, strained, portion of the sample, a smaller peak is visible at $\approx 1.45$~K, corresponding to the transition in the end portions.  This feature shows that temperature oscillations extend into the sample ends at these frequencies.
To avoid this, one has to work at the high end of the feasible range of frequencies. For this particular sample, a measurement frequency above $\sim1.5$~kHz was required.
Working at high frequencies with low enough power to avoid heating gives a very small signal, an r.m.s.\ thermocouple voltage of only $1 - 2$~nV. Therefore, the described low temperature passive amplification was employed to achieve an r.m.s.\ noise level of 20~pVHz$^{-1/2}$, ensuring a signal-to-noise ratio in excess of 50.

\subsection{Heat-capacity results on Sr$_2$RuO$_4$}

\begin{figure}[tb!]
\includegraphics[width=0.95\linewidth]{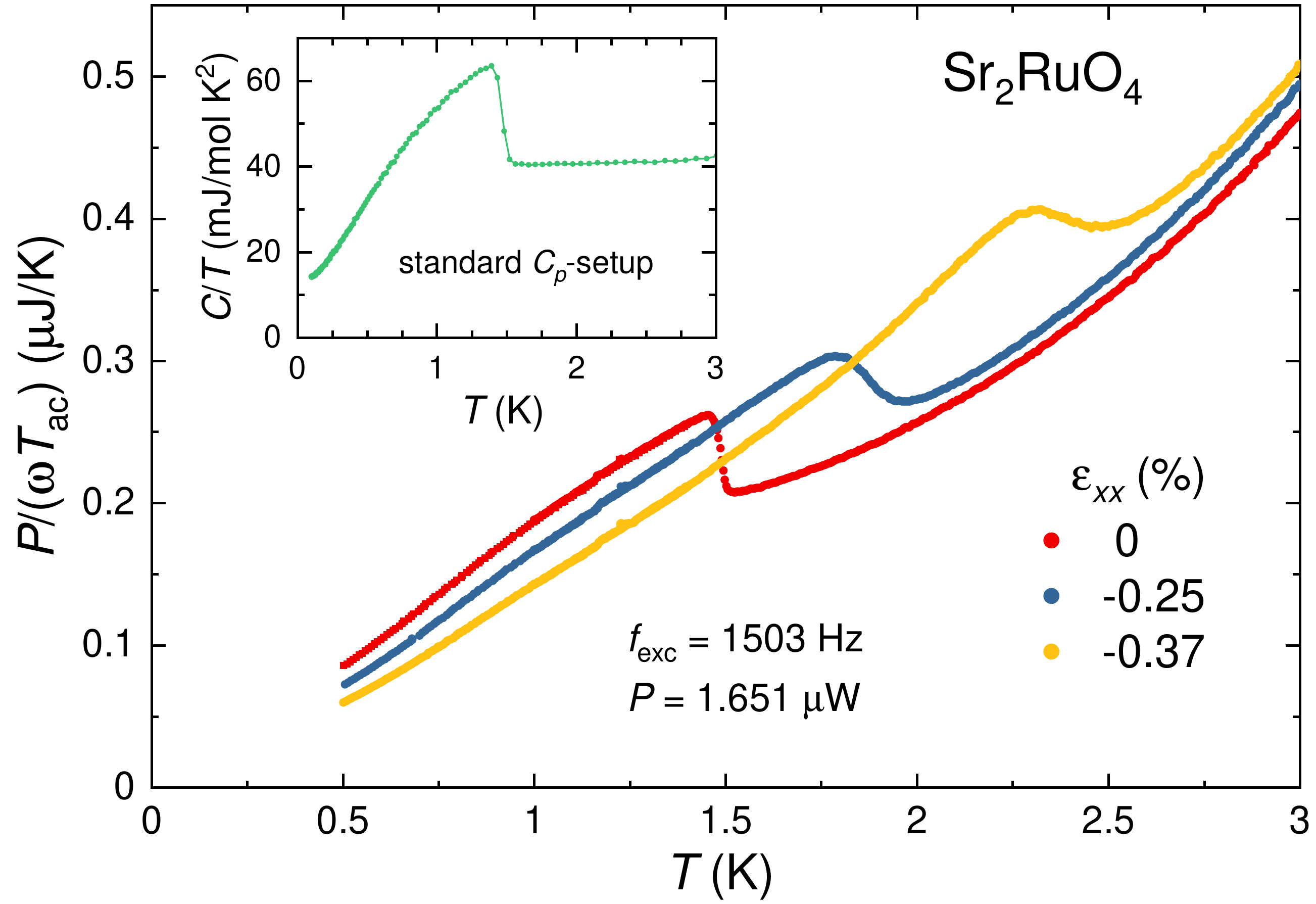}
\centering
\caption{Recorded signal $P/[\omega T_{AC}(T)]$ of Sr$_2$RuO$_4$ as function of temperature for three different strains $\varepsilon_{xx}=0$\%, $-0.25$\%, and $-0.37$\%. The inset shows a specific-heat experiment on a piece from the same crystal using a standard relaxation time method.
}
\label{HC_Sr2RuO4}
\end{figure}

Based on considerations outlined in the previous section we selected an excitation frequency of $f_{exc}=1503$~Hz to measure the heat capacity of Sr$_2$RuO$_4$. The results for three different strains $\varepsilon_{xx}=0$\%, $-0.25$\%, and $-0.37$\% are presented in Fig.\ \ref{HC_Sr2RuO4} as $P/[\omega T_{AC}(T)]$. Additionally the inset shows the results from a standard relaxation-type heat-capacity measurement from a piece of sample cut from the same crystal. It is qualitatively similar to the results in the uniaxial pressure cell at zero strain.

According to Eq.\ \ref{C_ac_prox} we find $C_{AC}(T)\approx P/[\omega T_{AC}(T)]$. However, this relation has to be taken with caution since the probed sample volume is not constant as function of temperature. We have selected $f_{exc}$ in order to probe the homogenously strained portion of the sample, but we have to notice that the thermal conductivity $\kappa$ of any studied material varies as function of temperature and strain, and as a consequence the probed sample volume also changes. To obtain absolute values of the volume specific heat $c_v(T)$ at a certain strain the temperature dependence of the thermal conductivity $\kappa(T)$ has to be known at that strain too.

\subsection{Determination of the volume specific heat}\label{Cv}

The conversion between the measured signal and the volume or molar specific heat is trivial in a conventional setup, because the volume (or mass) of the sample is constant. In our measurements, the probed sample volume varies since the thermal diffusion length $l_d$, which depends on thermal conductivity, specific heat, and frequency, changes as a function of temperature. Therefore, it is nontrivial to convert our data $P/[\omega T_{AC}(T)]$ to volume specific heat $c_v$. We start with an ideal case to demonstrate the relation between $P/[\omega T_{AC}(T)]$ and $c_v$ in case of our experimental setup.
Suppose that the heater contact is point-like in the center of a very narrow sample such that the heat flow is one-dimensional propagating in the left and right direction. The probed volume $V$ is equal to the cross-sectional area $A$ times twice the diffusion length $l_d$, which is a function of the angular frequency $\omega$, the volume specific heat $c_v$ and the thermal conductivity $\kappa$.
\begin{equation}
  l_d=\sqrt{\frac{2\kappa(T)}{\omega c_v(T)}}	
  \label{S1}
\end{equation}
$C_{AC}$ can be expressed as follows:
\begin{equation}\label{S2}
  C_{AC}=c_v \times V = c_v \times A\times l_d = \frac{2A}{\sqrt{\omega}} \sqrt{2\kappa(T) c_v(T)}.
\end{equation}
By using Eq.\ \ref{Cac} and \ref{S2} we finally obtain the volume specific heat $c_v$:
\begin{equation}\label{S3}
  c_{v}(T)=\left (\frac{P \times F(\omega)}{2A}\right )^2 \times \frac{1}{\omega\times 2\kappa(T)} \times \frac{1}{[T_{AC}(T)]^2}.
\end{equation}	
This exemplifies the reciprocal dependence of $c_v(T)$ on the thermal conductivity and the square of the temperature-oscillation amplitude in case of a simplified one dimensional model.

We further note that the excitation frequency in our current measurement is not too far away from the upper cut-off frequency, which describes the time scale for the heat propagating from the heater to the thermocouple. At this excitation frequency $F(\omega)<1$ and depends on temperature adding a further uncertainty on the determination of $c_v(T)$.

The validity of the Eqs.\ \ref{S2} and \ref{S3} is based on the above-mentioned assumptions that the heater contact is point-like and the heat flow is one-dimensional. In reality, both the sample width and the heater contact size are finite. This implies for the experimental setup to satisfy the assumptions of the examined model system, the exposed sample length ($l_s$) must be far longer than the heater length ($l_h$) and the sample width ($w_s$), $l_s \gg l_h,w_s$. Our present setup is already a good approximation to an ideal configuration but could in principle be further optimized.

In spite of the above caveats, we note that in some cases quantitative statements on the evolution of the specific heat on varying uniaxial pressure are possible based on the presently accessible data. For example, in superconductors, as in the case of Sr$_2$RuO$_4$, it is possible to obtain information on the evolution of the size of the superconducting transition anomaly with pressure, which is an important quantity characterizing superconductivity. In that case the thermal conductivity does not show any abrupt change across the transition and close to $T_c$
\begin{equation}\label{S4}
  \frac{c_{v}^s}{c_{v}^n}=\frac{\kappa_n}{\kappa_s} \times \left(\frac{T_{AC}^n}{T_{AC}^s}\right)^2 \approx \left(\frac{T_{AC}^n}{T_{AC}^s}\right)^2
\end{equation}	
with $\kappa_n\approx\kappa_s$.
The indices $s$ and $n$ indicate the corresponding values in the superconducting and in the normal state, respectively.

\section{CONCLUSION}

We have developed a new experimental setup using piezoelectric-driven uniaxial pressure cells for probing heat capacity at low temperatures. By optimizing our preparation and measuring processes we achieve an extremely high resolution and a high strain homogeneity in the probed sample volume.
The technique can be easily extended to different temperature regions. In addition to temperature sweeps, heat capacity can be recorded as function of applied pressure, and our apparatus is also fully compatible with work in magnetic fields.

\begin{acknowledgments}
We thank A.\ S.\ Gibbs, F.\ Jerzembeck, N.\ Kikugawa, Y.\ Maeno, D.\ A.\ Sokolov for providing and characterizing the samples and M.\ Brando and U.\ Stockert for experimental support.
\end{acknowledgments}

\section*{DATA AVAILABILITY}
The data that support the findings of this study are available from the corresponding author upon reasonable request.

\section*{REFERENCES}

\bibliography{Sr2RuO4}

\end{document}